\documentclass{ws-ijmpa}
\usepackage[super,compress]{cite}
\usepackage{graphicx}
\usepackage{amsbsy}

\def\mprp{\mbox{\tiny $\bot$}}
\def\mprl{\mbox{\tiny $\|$}}

\newcommand{\bs}{\boldsymbol} 
\newcommand{\I}{\mathrm{i}} % imaginary unit
\newcommand{\D}{\mathrm{d}} % differential
\newcommand{\E}{\mathrm{e}} % 2.71828...
\newcommand{\ee}{\varepsilon} 

\begin{document}
\markboth{A.\,V.~Kuznetsov, D.\,A.~Rumyantsev, V.\,N.~Savin}
{Creation of electron-positron pairs at excited Landau levels by neutrino \dots}
\catchline{}{}{}{}{}
%
%%%%%%%%%%%%%%%%%%%%%%%%%%%%%%%%%%%%%%%%%%%%%%%%%%%%%%%%%%%%%%%%%%%%

\title{
CREATION OF ELECTRON-POSITRON PAIRS AT EXCITED LANDAU LEVELS BY NEUTRINO IN A STRONG MAGNETIC FIELD
%Creation of electron-positron pairs at excited Landau levels by neutrino in a strong magnetic field
}

\author{A.\,V.~KUZNETSOV, D.\,A.~RUMYANTSEV}

\address{Division of Theoretical Physics, Department of Physics,\\
Yaroslavl State P.\,G.~Demidov University, Sovietskaya 14,\\
150000 Yaroslavl, Russian Federation\\
avkuzn@uniyar.ac.ru, rda@uniyar.ac.ru}

\author{V.\,N.~SAVIN}

\address{Division of Physics, \\
A.\,F.~Mozhaiskiy Space Military Academy, Yaroslavl Branch,\\
Moskovskiy Prosp. 28, 150001 Yaroslavl, Russian Federation\\
vs\_post07@mail.ru}

\maketitle

\begin{history}
\received{Day Month Year}
\revised{Day Month Year}
\end{history}

\begin{abstract}
The process of neutrino production of electron positron pairs in a magnetic field of arbitrary strength, where 
electrons and positrons can be created in the states corresponding to excited Landau levels, is analysed. 
The mean value of the neutrino energy loss due to the process 
$\nu \to \nu e^- e^+$ is calculated. The result can be applied for calculating the efficiency of the electron-positron plasma production by neutrinos 
in the conditions of the Kerr black hole accretion disc considered by experts as the most possible source of a short 
cosmological gamma burst. The presented research can be also 
useful for further development of the calculation technic for an analysis of quantum processes 
in external active medium, and in part in the conditions of moderately strong magnetic field, when taking 
account of the ground Landau level appears to be insufficient.

\end{abstract}

\ccode{PACS numbers: 13.15.+g, 95.30.Cq}

%\tableofcontents

%%%%%%%%%%%%%%%%%%%%%%%%%%%%%%%%%%%%%%%%%%%%%%%%%%%%%%%%%%%%%%%%%%%%%%%%%%%%%%%%
\section{Introduction}
\label{sec:Introduction}
%%%%%%%%%%%%%%%%%%%%%%%%%%%%%%%%%%%%%%%%%%%%%%%%%%%%%%%%%%%%%%%%%%%%%%%%%%%%%%%%

An intense electromagnetic field makes possible the processes which are forbidden in a vacuum such as 
the neutrino production of an electron--positron pair $\nu \to \nu e^- e^+$. 
The list of papers devoted to an analysis of this process and the collection of the results obtained 
could be found e.g. in Ref.~\refcite{KM_Book_2013}. 
In most cases, calculations of this kind were made either in the crossed field approximation, 
or in the limit of a superstrong field much greater than the critical value of 
$B_e = m_e^2/e \simeq 4.41\times 10^{13}$~G, 
when the electrons and positrons are born
in states corresponding to the ground Landau level. 
However, there are physical situations of the so-called moderately strong magnetic 
field,\footnote{We use natural units
$c = \hbar = k_{\rm{B}} = 1$, $m_e$ is the electron mass, and $e$ is the elementary
charge.} 
$p_\perp^2 \gtrsim e B \gg m_e^2$, when electrons and positrons mainly occupy the ground
Landau level, however, a noticeable fraction may be produced at the next levels. 

The indicated hierarchy of physical parameters corresponds to the conditions of the Kerr black hole accretion disk, regarded by experts as the most likely source of a short cosmological gamma-ray burst. 
The disc is a source of copious neutrinos and anti-neutrinos, which partially annihilate above the disc
and turn into $e^{\mp}$ pairs, $\nu \bar\nu \to e^- e^+$. This process was proposed and investigated in many details~\cite{Berezinsky:1987,Eichler:1989,Ruffert:1997,Asano:2000,Asano:2001,Birkl:2007} 
as a possible mechanism for creating relativistic, $e^{\mp}$-dominated jets that could power observed 
gamma-ray bursts. 
In Ref.~\refcite{Beloborodov:2011}, in addition to $\nu \bar\nu$ annihilation, the contribution
of the magnetic field-induced process $\nu \to \nu e^- e^+$ to the neutrino energy deposition rate around
the black hole was also included. 

However, 
in calculations of the efficiency of the electron-positron plasma production by neutrino through 
the process $\nu \to \nu e^- e^+$ in these physical conditions, it should be kept in mind that approximations 
of both the crossed and superstrong field have a limited applicability here. 
We know a limited number of papers~\cite{Bezchastnov:1996,Mikheev:2000,Mikheev:2003,Dicus:2007}, 
where the probability of neutrino-electron processes was investigated,
as the sum over the Landau levels of electrons (positrons).
In the papers, Refs.~\refcite{Bezchastnov:1996}-\refcite{Mikheev:2003},
only the neutrino-electron scattering channel in a dense magnetized plasma was studied, 
which was the crossed process to the considered here neutrino creation of electron-positron pairs.
In the paper Ref.~\refcite{Dicus:2007}, also devoted to the study of the process $\nu \to \nu e^- e^+$, the analytical
calculations were presented in a rather cumbersome form, caused by the choice of solutions of the Dirac equation.
The final results for the process probability were obtained by numerical calculations for some set of Landau levels 
occupied by electrons and positrons. 
In astrophysical applications, there exists probably more interesting value than the process probability, 
namely, the mean value of the neutrino energy loss, caused by the influence of an external magnetic field. 
Thus, the aim of this paper is the study of the process $\nu \to \nu e^- e^+$ in the physical conditions of the moderately strong 
magnetic field, where the electrons and positrons would be born in the states corresponding to the excited
Landau levels, and the theoretical description would contain a relatively simple analytical formulas 
both for the process probability and the mean value of the neutrino energy loss, for a wide range of Landau levels. 
In Sec.~\ref{sec:Solutions}, the exact solutions of the Dirac equation
for the electron and positron in a magnetic field are presented, being the eigenfunctions 
of the covariant operator of magnetic polarization. 
Sec.~\ref{sec:amplitudes} is devoted to the calculation of the partial polarization amplitudes of the process, 
both in general terms and in special cases when the results are described by rather simple analytical formulas.
In Sec.~\ref{sec:probability}, the probabilities are obtained for the main polarization channels, 
presented in the form of double and triple integrals. 
The final results obtained by numerical calculations for all channels considered in 
Ref.~\refcite{Dicus:2007} are in a good agreement with the results of that paper.
Sec.~\ref{sec:losses} is devoted to the calculation of the neutrino energy losses caused by
the process $\nu \to \nu e^- e^+$ in a moderately strong magnetic field i.e. in the conditions 
of the Kerr black hole accretion disk. Possible astrophysical applications are discussed. 

%%%%%%%%%%%%%%%%%%%%%%%%%%%%%%%%%%%%%%%%%%%%%%%%%%%%%%%%%%%%%%%%%%%%%%%%%%%%%%%%
\section{Solutions of the Dirac equation for an electron \newline in a magnetic field}	
\label{sec:Solutions}
%%%%%%%%%%%%%%%%%%%%%%%%%%%%%%%%%%%%%%%%%%%%%%%%%%%%%%%%%%%%%%%%%%%%%%%%%%%%%%%%

There exist several descriptions of the procedure of obtaining the electron wave functions in the presence 
of an external magnetic field by solving the Dirac equation, see
e.g. Refs.~\refcite{Johnson:1949}--\refcite{Balantsev:2011} and 
also Refs.~\refcite{KM_Book_2013,KM_Book_2003}. 
In the most cases, the solutions are presented in the form with the upper two components 
of the bispinor corresponding to the electron states with the spin projections 1/2 and -1/2
on the magnetic field direction.
Here, we have found it more convenient to use another representation of the electron wave functions. 

In Ref.~\refcite{Sokolov:1968}, an operator was introduced which was called the generalized spin 
tensor of the third rank. In modern standard notations, the operator takes the 
form\footnote{It should be noted that in Ref.~\refcite{Sokolov:1968}, the covariant bilinear forms 
were constructed of Dirac matrices by inserting them not between bispinors $\bar\psi$ and $\psi$ 
as accepted in modern literature,~\cite{Peskin:1995} but between bispinors $\psi^{\dagger}$ and $\psi$.}
\begin{eqnarray}
{\rm F}_{\mu \nu \lambda} = - \frac{\I}{2} \left( P_\lambda \gamma_0 \sigma_{\mu \nu} 
+ \gamma_0 \sigma_{\mu \nu} P_\lambda \right),
\label{eq:Fgen}
\end{eqnarray}
where $\sigma_{\mu \nu} = (\gamma_\mu \gamma_\nu - \gamma_\nu \gamma_\mu)/2$, and 
$P^\lambda = \I \partial^\lambda + e \, A^\lambda = \left( \I \partial_0 + e \, A_0 \,, 
- \I {\bs \nabla} + e {\bs A} \right)$ is the generalized four-momentum operator with $A^\lambda$ 
being the four-potential of an external magnetic field. 
Taking the component ${\rm F}_{\mu \nu 0}$ of the operator~(\ref{eq:Fgen}) and taking into 
account that in the Schr\"odinger form of the Dirac equation one has $\I \partial_0 = H$, where 
$H = \gamma_0 \left( {\bs \gamma} {\bs P} \right) + m_e \, \gamma_0 - e A_0$ 
is the Dirac Hamiltonian, one can construct the vector operator
\begin{eqnarray}
{\mu}_i = - \frac{1}{2} \, \varepsilon_{ijk} \, {\rm F}_{jk0} \,, 
\label{eq:mu_i}
\end{eqnarray}
where $\varepsilon_{ijk}$ is the Levi-Civita symbol. This is the magnetic moment operator,~\cite{Sokolov:1968,Melrose:1983} which can be presented in the form
\begin{eqnarray}
{\bs \mu} = m_e {\bs \Sigma} - \I \gamma_0 \gamma_5 [{\bs \Sigma} \times \hat{\bs P}] \,. 
\label{eq:mu_vec}
\end{eqnarray}
It is straightforward to show that the components of the operator~(\ref{eq:mu_vec}) 
commute with the Hamiltonian, i.e. $H$ and ${\mu}_z$ have common eigenfunctions. 
In the non-relativistic limit, the operator~(\ref{eq:mu_vec}) is transformed to the ordinary Pauli magnetic moment
operator, thus having an obvious physical interpretation. 

It appears to be convenient to use the electron wave functions as the 
eigenstates of the operator ${\mu}_z$~\cite{Sokolov:1968,Melrose:1983} 
\begin{eqnarray}
{\mu}_z = m_e \Sigma_z - \I \gamma_0 \gamma_5 [{\bs \Sigma} \times {\bs P}]_z \,, 
\label{eq:mu_z}
\end{eqnarray}
where ${\bs P} =  - \I {\bs \nabla} + e {\bs A}$. 
We take the frame where the field is directed 
along the $z$ axis, and the Landau gauge where the four-potential is: $A^\lambda = (0, 0, x B, 0)$. 

In this approach, the electron wave functions have the form
\begin{eqnarray}
\label{eq:psie}
\Psi^s_{p,n}(X) = \frac{\E^{-\I(\ee_{n} t - p_y y - p_z z)}\; u^s_{n} (\xi)}
{\sqrt{4 \ee_{n}M_n (\ee_{n} + M_n)(M_n + m_e) L_y L_z}} \, ,  
\end{eqnarray}
where
\begin{eqnarray}
\label{eq:E_n,M_n}
\ee_n = \sqrt{M_n^2 + p_z^2}\, , \quad  M_n = \sqrt{m_e^2 + 2 \beta n}\, ,\quad  \beta = e B \,,
\quad
\xi = \sqrt{\beta} \left (x + \frac{p_y}{\beta} \right ) .
\end{eqnarray}
Note that the value $p_z$ in Eq.~(\ref{eq:psie}) is a conserved
component of the electron momentum along the $z$ axis, 
i.e. along the field, while the value $ p_y $ is the generalized momentum, 
which determines the position of a center of the Gaussian packet along 
the $x$ axis.

The functions $\Psi^s_{p,n}(X)$ satisfy the equation:
\begin{eqnarray}
\label{eq:mu_z_Eq}
\hat{\mu}_z \,\Psi^s_{p,n}(X) = s \, M_n \, \Psi^s_{p,n}(X) \, , \quad s = \pm 1\,.
\end{eqnarray}
The bispinors $u^s_{n} (\xi)$ in Eq.~(\ref{eq:psie}) take the form:
\begin{eqnarray}
\label{eq:U--}
&&u^{-}_{n} (\xi) = \left ( 
\begin{array}{c}
-\I\sqrt{2\beta n} \, p_z V_{n-1} (\xi)\\[2mm]
(\ee_n + M_n)(M_n + m_e) V_n (\xi)\\[2mm]
-\I\sqrt{2\beta n} (\ee_n + M_n) V_{n-1} (\xi)\\[2mm]
-p_z (M_n + m_e) V_n (\xi)
\end{array}
\right )  ,   
\\ [3mm]
\label{eq:U+-}
&&u^{+}_{n} (\xi) = \left ( 
\begin{array}{c}
(\ee_n + M_n) (M_n + m_e) V_{n-1} (\xi)\\[2mm]
-\I\sqrt{2\beta n} \, p_z V_n (\xi)\\[2mm]
p_z (M_n + m_e) V_{n-1} (\xi)\\[2mm]
\I \sqrt{2 \beta n} (\ee_n + M_n) V_n (\xi)
\end{array}
\right )\! . 
\end{eqnarray}
Here, $V_n(\xi) \, (n = 0,1,2, \dots)$ 
are the normalized harmonic oscillator functions, which are expressed in terms of 
Hermite polynomials $H_n(\xi)$ \cite{Gradshtein}:
\begin{eqnarray}
\label{eq:V_n}
V_n (\xi) = \frac{\beta^{1/4}\E^{-\xi^2/2}}{\sqrt{2^n n! \sqrt{\pi}}} \, H_n(\xi)\, .
\end{eqnarray}

The wave function of an electron with negative energy that corresponds
to a positron in a final state, with positive energy $\ee'_{\ell}$ 
and the momentum components $p'_y, \, p'_z$, in the presence of external 
magnetic field, which also satisfies the equation~(\ref{eq:mu_z_Eq}), 
can be written as:
\begin{eqnarray}
\label{eq:psie+}
\Psi^{s'}_{p',\ell}(X) = \frac{\E^{\I(\ee'_{\ell} t - p'_y y - p'_z z)}\; v^{s'}_{\ell} (\xi')}
{\sqrt{4 \ee'_{\ell} M_{\ell} (\ee'_{\ell} + M_{\ell})(M_{\ell} + m_e) L_y L_z}} \, ,  
\end{eqnarray}
\begin{eqnarray}
\label{eq:E_l,M_l}
\ee'_{\ell} = \sqrt{M_{\ell}^2 + {p'}_z^2}\, , \quad  M_{\ell} = \sqrt{m_e^2 + 2 \beta {\ell}}\, ,
\quad  
\xi' = \sqrt{\beta} \left (x - \frac{p'_y}{\beta} \right ) .
\end{eqnarray}
The bispinors $v^{s'}_{\ell} (\xi')$ in Eq.~(\ref{eq:psie+}) take the form:
\begin{eqnarray}
\label{eq:U-+}
v^{-}_{\ell} (\xi') = \left ( 
\begin{array}{c}
\I\sqrt{2\beta {\ell}} \, (\ee'_{\ell} + M_{\ell}) V_{{\ell}-1} (\xi')\\[2mm]
- p'_z (M_{\ell} + m_e) V_{\ell} (\xi')\\[2mm]
\I\sqrt{2\beta {\ell}} \, p'_z \, V_{{\ell}-1} (\xi')\\[2mm]
(\ee'_{\ell} + M_{\ell}) (M_{\ell} + m_e) V_{\ell} (\xi')
\end{array}
\right )  ,   
\\ [3mm]
\label{eq:U++}
v^{+}_{\ell} (\xi') = \left ( 
\begin{array}{c}
p'_z (M_{\ell} + m_e) V_{{\ell}-1} (\xi')\\[2mm]
-\I\sqrt{2\beta {\ell}} \, (\ee'_{\ell} + M_{\ell}) V_{\ell} (\xi')\\[2mm]
(\ee'_{\ell} + M_{\ell}) (M_{\ell} + m_e) V_{{\ell}-1} (\xi')\\[2mm]
\I \sqrt{2 \beta {\ell}} \, p'_z \, V_{\ell} (\xi')
\end{array}
\right )\! . 
\end{eqnarray}

Obvious advantages of such a choice of the solutions of the Dirac equation will become apparent 
from the subsequent analysis.

%%%%%%%%%%%%%%%%%%%%%%%%%%%%%%%%%%%%%%%%%%%%%%%%%%%%%%%%%%%%%%%%%%%%%%%%%%%%%%%%
\section{Partial polarization amplitudes}	
\label{sec:amplitudes}
%%%%%%%%%%%%%%%%%%%%%%%%%%%%%%%%%%%%%%%%%%%%%%%%%%%%%%%%%%%%%%%%%%%%%%%%%%%%%%%%

We use the standard calculation technics, see e.g. Ref.~\refcite{KM_Book_2013}. 
The effective local Lagrangian of the process can be written in the form
\begin{equation}
{\cal L} \, = \, - \frac{G_{\mathrm{F}}}{\sqrt 2} 
\big [ \bar e \gamma_{\alpha} (C_V - C_A \gamma_5) e \big ] \,
\big [ \bar \nu \gamma^{\alpha} (1 - \gamma_5) \nu \big ] \,,
\label{eq:L}
\end{equation}
where the electron field operators are constructed on a base of the above-mentioned 
solutions of the Dirac equation. 
The constants $C_V$ and $C_A$ for different neutrino types are:
\begin{eqnarray}
&&C_V^{(e)} = + \frac{1}{2} + 2 \sin^2 \theta_{\rm W} \,, \qquad C_A^{(e)} = + \frac{1}{2} \,,
\nonumber\\%[2mm]
&&C_V^{(\mu,\tau)} = - \frac{1}{2} + 2 \sin^2 \theta_{\rm W} \,, \qquad C_A^{(\mu,\tau)} = - \frac{1}{2} \,,
\label{eq:CVCA}
\end{eqnarray}
where $\theta_\mathrm{W}$ is the Weinberg angle.
The conditions of applicability of the effective Lagrangian~(\ref{eq:L}) should be specified. 
First, it is the condition of relatively small momentum transfers, $|q^2| \ll m_W^2\,$, 
where $m_W$ is the $W$ boson mass. And second, the condition that additionally arises 
in an external magnetic field, is $e B \ll m_W^2\,$. Both of these conditions are obviously 
satisfied in the considered physical situation.

Calculation of the ${\cal S}$ matrix element of the process
$\nu \to \nu e^-_{(n)} e^+_{(\ell)}$, when the electron and the positron are created 
in the $n$th and $\ell$th Landau levels, is more complicated computational task than in 
the case of the ground Landau level.~\cite{KM_Book_2013}
The following integrals appear in the calculations: 
\begin{eqnarray}
\nonumber
&&\frac{1}{\sqrt{\pi}}\int \D Z \, \E^{-Z^2} 
H_n \! \left (Z + \frac{q_y + \I q_x}{2\sqrt{\beta}} \right ) 
H_{\ell} \! \left (Z - \frac{q_y - \I q_x}{2\sqrt{\beta}} \right ) 
\\[3mm]
&&= 2^{(n+\ell)/2} \sqrt{n ! \, \ell !} 
\left (\frac{q_y + \I q_x}{\sqrt{q^2_{\mprp}}} \right )^{n-\ell} 
\E^{{q^2_{\mprp}}/{(4\beta)}} \,
{\cal I}_{n, \ell} \! \left (\frac{q^{2}_{\mprp}}{2 \beta} \right ) \, , 
\label{eq:Ilnnl}
\end{eqnarray}
where, for $n \geqslant \ell$
\begin{eqnarray}
\nonumber
&&{\cal I}_{n, \ell} (x) = \sqrt{\frac{\ell !}{n !}} \; \E^{-x/2} x^{(n-\ell)/2} L_\ell^{n-\ell} (x) \, ,
\\
&&{\cal I}_{\ell, n} (x) = (-1)^{n-\ell} {\cal I}_{n, \ell} (x) \, ,
\label{eq:Inl}
\end{eqnarray}
and $L^k_n (x)$ are the generalized Laguerre polynomials~\cite{Gradshtein}.

Hereafter we use the following notations: $\varphi_{\alpha \beta} =  F_{\alpha \beta} /B$ 
is the dimensionless tensor of the external magnetic field,  
${\tilde \varphi}_{\alpha \beta} = \frac{1}{2} \, \varepsilon_{\alpha \beta
\mu \nu} \varphi^{\mu \nu}$ is the dual dimensionless tensor;
the dimensionless tensors
$\Lambda_{\alpha \beta} = (\varphi \varphi)_{\alpha \beta}$,\,  
$\widetilde \Lambda_{\alpha \beta} = 
(\tilde \varphi \tilde \varphi)_{\alpha \beta}$ are connected by the 
relation 
$\widetilde \Lambda_{\alpha \beta} - \Lambda_{\alpha \beta} = g_{\alpha \beta}$. 
The tensor indices of four-vectors and tensors standing inside the 
parentheses are contracted consecutively, e.g.:
$(\varphi \varphi p)_\alpha = \varphi_{\alpha 
\beta} \varphi^{\beta \lambda} p_{\lambda}$. 
The four-vectors with the indices
$\bot$ and $\parallel$ belong to the 
Euclidean \{1, 2\}-subspace\index{Euclidean \{1, 2\}-subspace}  
and the Minkowski \{0, 3\}-subspace\index{Minkowski \{0, 3\}-subspace}, 
correspondingly (we remind that the magnetic field is directed along the 3d axis), 
then $p_\bot^\mu = (0, p_1, p_2, 0)$, $p_\|^\mu = (p_0, 0, 0, p_3)$, and 
$\Lambda_{\alpha \beta} = \textrm{diag}(0, 1, 1, 0)$, 
$\widetilde \Lambda_{\alpha \beta} = \textrm{diag}(1, 0, 0, -1)$.
For arbitrary four-vectors $p^\mu$, $q^\mu$ one has
$(pq)_\bot = (p \Lambda q) =  p_1 q_1 + p_2 q_2$, 
$(pq)_\| = (p \widetilde \Lambda q) = p_0 q_0 - p_3 q_3$, and
$(pq) = (pq)_\| - (pq)_\bot$. 

An invariant amplitude of the process $\nu \to \nu e^-_{(n)} e^+_{(\ell)}$ 
is extracted by the standard way from the ${\cal S}$ matrix element:
\begin{eqnarray}
{\cal S}_{n \ell}^{s s'} = \frac{\I (2 \pi)^3 \, \delta^{(3)}(p + p' - q)}
{\sqrt{2 E V\,2 E' V\, 
2\varepsilon_n L_y L_z\, 2\varepsilon'_{\ell} L_y L_z} } \, {\cal M}_{n \ell}^{s s'}\, , 
\label{eq:Smat} 
\end{eqnarray}
where $\delta^{(3)}(p + p' - q) = \delta(\varepsilon_n + \varepsilon'_{\ell} - q_0) \,\delta(p_y + p_y' - q_y) \,
\delta(p_z + p_z' - q_z)$, 
$q = P - P' = p + p'$ is the change of the four-vector of the neutrino 
momentum equal to the four-momentum of the $e^- e^+$ pair, and $V = L_x L_y L_z$ is the total volume 
of the interaction region.  

Constructing the ${\cal S}$ matrix element with the effective Lagrangian~(\ref{eq:L}) 
and extracting the invariant amplitude according to Eq.~(\ref{eq:Smat}), we calculate 
the four invariant polarization amplitudes ${\cal M}_{n \ell}^{s s'}$ where $s, s' = \pm 1$,
by direct multiplication of the bispinors (\ref{eq:U--}), (\ref{eq:U+-}), (\ref{eq:U-+}), (\ref{eq:U++}).
The amplitudes ${\cal M}^{--}_{n \ell}$ and ${\cal M}^{++}_{n \ell}$ can be written with the single formula:
\begin{eqnarray}
&&{\cal M}^{\mp \mp}_{n \ell} = \pm \, \eta  
\, \frac{G_{\mathrm{F}}}
{2 \, \sqrt{2} } \, 
\bigg \lbrace 
\sqrt{\left( 1 \pm \frac{m_e}{M_{n}} \right)
\left( 1 \pm \frac{m_e}{M_{\ell}} \right)}
\,
{\cal I}_{n, \ell} \;
[ C_V \,(j \, {\cal K}_2) + C_A \,(j \, {\cal K}_1)] 
\nonumber\\[2mm]
&&+ \,
\sqrt{\left( 1 \mp \frac{m_e}{M_{n}} \right)
\left( 1 \mp \frac{m_e}{M_{\ell}} \right)} \;
{\cal I}_{n-1, \ell-1}
\,
[ C_V \,(j \, {\cal K}_2) - C_A \,(j \, {\cal K}_1)]  
\nonumber\\[2mm]
&&+ \,
\sqrt{\left( 1 \pm \frac{m_e}{M_{n}} \right)
\left( 1 \mp \frac{m_e}{M_{\ell}} \right)} \;
{\cal I}_{n, \ell-1}
\,
\left( C_V \, {\cal K}_4 + C_A \, {\cal K}_3 \right) 
\frac{(j \Lambda q) + \I \, (j \varphi q)}{\sqrt{q^2_{\mprp}}}
\nonumber\\[2mm]
&&+ \,
\sqrt{\left( 1 \mp \frac{m_e}{M_{n}} \right)
\left( 1 \pm \frac{m_e}{M_{\ell}} \right)} \;
{\cal I}_{n-1, \ell}
\,
\left( C_V \, {\cal K}_4 - C_A \, {\cal K}_3 \right) 
\frac{(j \Lambda q) - \I \, (j \varphi q)}{\sqrt{q^2_{\mprp}}}
\bigg \rbrace ,
\label{eq:M_--_nl}
\end{eqnarray}
and similarly the amplitudes ${\cal M}^{+-}_{n \ell}$ and ${\cal M}^{-+}_{n \ell}$:
\begin{eqnarray}
&&{\cal M}^{\pm \mp}_{n \ell} = \I \, \eta  
\, \frac{G_{\mathrm{F}}}
{2 \, \sqrt{2} } \, 
\bigg \lbrace 
\sqrt{\left( 1 \mp \frac{m_e}{M_{n}} \right)
\left( 1 \pm \frac{m_e}{M_{\ell}} \right)}
\,
{\cal I}_{n, \ell} \;
[ C_V \,(j \, {\cal K}_1) + C_A \,(j \, {\cal K}_2)] 
\nonumber\\[2mm]
&&- \,
\sqrt{\left( 1 \pm \frac{m_e}{M_{n}} \right)
\left( 1 \mp \frac{m_e}{M_{\ell}} \right)} \;
{\cal I}_{n-1, \ell-1}
\,
[ C_V \,(j \, {\cal K}_1) - C_A \,(j \, {\cal K}_2)]  
\nonumber\\[2mm]
&&+ \,
\sqrt{\left( 1 \mp \frac{m_e}{M_{n}} \right)
\left( 1 \mp \frac{m_e}{M_{\ell}} \right)} \;
{\cal I}_{n, \ell-1}
\,
\left( C_V \, {\cal K}_3 + C_A \, {\cal K}_4 \right) 
\frac{(j \Lambda q) + \I \, (j \varphi q)}{\sqrt{q^2_{\mprp}}}
\nonumber\\[2mm]
&&- \,
\sqrt{\left( 1 \pm \frac{m_e}{M_{n}} \right)
\left( 1 \pm \frac{m_e}{M_{\ell}} \right)} \;
{\cal I}_{n-1, \ell}
\,
\left( C_V \, {\cal K}_3 - C_A \, {\cal K}_4 \right) 
\frac{(j \Lambda q) - \I \, (j \varphi q)}{\sqrt{q^2_{\mprp}}}
\bigg \rbrace ,
\label{eq:M_+-_nl}
\end{eqnarray}
Here, $\eta$ is an inessential phase factor, 
$j^\alpha$ is the Fourier transform of the neutrino current,
and we have omitted the argument $(q^{2}_{\mprp}/{2 \beta})$ of the functions 
${\cal I}$ defined by Eqs.~(\ref{eq:Inl}). 
The following auxiliary covariants are inserted in Eqs.~(\ref{eq:M_--_nl})--(\ref{eq:M_+-_nl}),
the 4-vectors in the $\{0,3\}$ subspace:
\begin{eqnarray}
\label{eq:K1}
&&{\cal K}_1^\alpha = \sqrt{2} \, \frac{M_n (\widetilde \Lambda p^{\, \prime})^\alpha + 
M_{\ell} (\widetilde \Lambda p)^\alpha}{\sqrt{(p\widetilde \Lambda p^{\, \prime}) + 
M_n M_{\ell}}} \,,
\\
\label{eq:K2}
&&{\cal K}_2^\alpha = \sqrt{2} \, \frac{M_n (\widetilde \varphi p^{\, \prime})^\alpha + 
M_{\ell} (\widetilde \varphi p)^\alpha}{\sqrt{(p\widetilde \Lambda p^{\, \prime}) + 
M_n M_{\ell}}} \,,
\end{eqnarray}
and the invariants:
\begin{eqnarray}
\label{eq:K3}
&&{\cal K}_{3} = \sqrt{2\left [(p\widetilde \Lambda p^{\, \prime}) + 
M_n M_{\ell} \right]} \, ,  
\\
\label{eq:K4}
&&{\cal K}_4 = 
- \frac{\sqrt{2} \; (p\widetilde \varphi p^{\, \prime})}
{\sqrt{(p\widetilde \Lambda p^{\, \prime}) + M_n M_{\ell}}} \, .
\end{eqnarray}

We emphasize the remarkable property of the partial amplitudes~(\ref{eq:M_--_nl})--(\ref{eq:M_+-_nl}),
corresponding to different polarization states of the electrons and positrons, namely, 
their manifestly relativistic invariant form. On the contrary, the amplitudes obtained with using the solutions for 
a fixed direction of the spin, do not have Lorentz invariant structure. Only the amplitudes squared, 
summed over the electron and positron polarization states, are manifestly Lorentz-invariant.
Thus, our approach is an alternative to the method where the amplitudes squared are calculated,
summed over the fermion polarization states, with using the fermion density matrices,
see e.g. Refs.~\refcite{Andreev:2010,Gvozdev:2012}.

%%%%%%%%%%%%%%%%%%%%%%%%%%%%%%%%%%%%%%%%%%%%%%%%%%%%%%%%%%%%%%%%%%%%%%%%%%%%%%%%
\section{The probability of the process $\nu \to \nu e^- e^+$}	
\label{sec:probability}
%%%%%%%%%%%%%%%%%%%%%%%%%%%%%%%%%%%%%%%%%%%%%%%%%%%%%%%%%%%%%%%%%%%%%%%%%%%%%%%%

The total probability of the process $\nu \to \nu e^-_{(n)} e^+_{(\ell)}$
is, in a general case, the sum of the probabilities of the four polarization channels:
\begin{equation}
\label{eq:Wtot}
W_{n \ell} = W^{--}_{n \ell} + W^{-+}_{n \ell} + W^{+-}_{n \ell} + W^{++}_{n \ell} \, .
\end{equation}
For each of the channels, the differential probability over the final neutrino momentum 
per unit time can be written as
\begin{equation}
\D W^{s s'}_{n \ell} \, = \, \frac{1}{\cal T}\, 
\frac{\D^3 P'\,V}{(2 \pi)^3} \; 
\int \, |{\cal S}_{n \ell}^{s s'}|^2 \, 
\D \Gamma_{e^-} \;
\D \Gamma_{e^+} \,,
\label{eq:dw1} 
\end{equation}
where $\cal T$ is the total interaction time, and the elements of 
the phase volume are introduced for the electron and the positron:
\begin{equation}
\D \Gamma_{e^-} = \frac{\D^2 p\,L_y L_z}{(2 \pi)^2}, \quad
\D \Gamma_{e^+} = \frac{\D^2 p'\,L_y L_z}{(2 \pi)^2} \, .
\label{eq:d_n} 
\end{equation}
Given Eq.~(\ref{eq:Smat}), integration over the momenta of the electron and positron is reduced to one
nontrivial integral:
\begin{eqnarray}
\D W^{s s'}_{n \ell} = 
\frac{\beta \, \D^3 P'}{(2 \pi)^4 16 E E'} \, 
\int \, 
\frac{\D p_z}{\varepsilon_n \, \varepsilon'_{\ell}} \, \delta(\varepsilon_n + \varepsilon'_{\ell} - q_0) \,
|{\cal M}_{n \ell}^{s s'}|^2 \, .
\label{eq:dw2} 
\end{eqnarray}
After integration with $\delta$-functions, the covariants~(\ref{eq:K1})--(\ref{eq:K4}) can
be written in the form
\begin{eqnarray}
{\cal K}_1^\alpha &=& \frac{1}{q_{\mprl}^2}  
\left[ (M_{n} + M_{\ell}) \, {\cal K}_3 \, (\widetilde \Lambda q)^\alpha -
(M_{n} - M_{\ell}) \, {\cal K}_4 \, (\widetilde \varphi q)^\alpha \right] \,,
\label{eq:K1i}\\[2mm]
{\cal K}_2^\alpha &=& \frac{1}{q_{\mprl}^2}  
\left[ (M_{n} + M_{\ell}) \, {\cal K}_3 \, (\widetilde \varphi q)^\alpha -
(M_{n} - M_{\ell}) \, {\cal K}_4 \, (\widetilde \Lambda q)^\alpha \right] \,,
\label{eq:K2i}\\[2mm]
{\cal K}_3 &=& \sqrt{q_{\mprl}^2 - (M_{n} - M_{\ell})^2} \, ,  
\label{eq:K3i}\\[2mm]
{\cal K}_4 &=& \zeta \, \sqrt{q_{\mprl}^2 - (M_{n} + M_{\ell})^2} \, .
\label{eq:K4i}
\end{eqnarray}
Here, $\zeta = \pm 1$ is the sign factor associated with the two roots of the equation 
$\sqrt{{p}_z^2 + M_n^2} + \sqrt{(q_z - {p}_z)^2 + M_{\ell}^2} - q_0 = 0$, 
corresponding to the zeros of the $\delta$ function argument in Eq.~(\ref{eq:dw2}). 
In the frame where $q_z = 0$, $\zeta$ is the sign of the $p_z$ component, 
which is not fixed by the equation. 

From the analysis of the solvability of that equation, the condition arises: 
\begin{equation}
q_{\mprl}^2 \geqslant (M_{n} + M_{\ell})^2 \, ,
\label{eq:cond} 
\end{equation}
which determines the range of integration over the final neutrino momentum.

In turn, the condition~(\ref{eq:cond}) can be satisfied when the energy of the initial
neutrino exceeds a certain threshold value.
In the reference frame, hereafter called $K$, where the momentum of the initial neutrino directed at an angle
$\theta$ to the magnetic field, the threshold energy is given by:
\begin{equation}
E \, \sin \theta \geqslant M_{n} + M_{\ell} \, .
\label{eq:condE} 
\end{equation}

\begin{figure}[!t] 
\centering 
\includegraphics*[width=\textwidth]{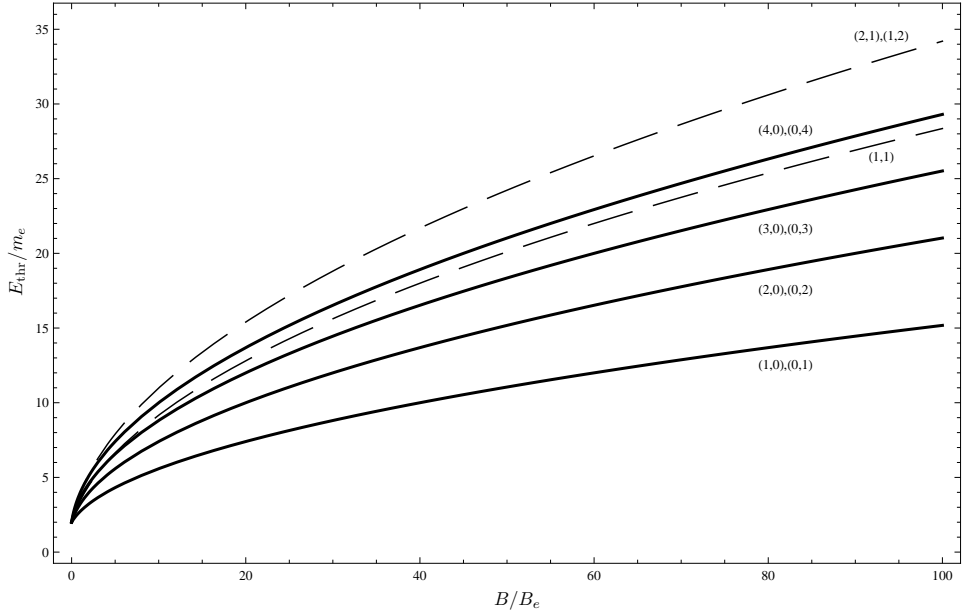}
\caption{Dependence of the threshold energy $E_{\rm{thr}}$ for the process $\nu \to \nu e^-_{(n)} e^+_{(\ell)}$ 
at $\theta = \pi/2$ on the magnetic field strength for several final states $(n, \ell)$, where the electron 
and the positron are created in the lower excited Landau levels.}
\label{fig:threshold}
\end{figure}

Fig.~\ref{fig:threshold} shows the dependence of the threshold energy $E_{\rm{thr}}$ at $\theta = \pi/2$
on the strength of the magnetic field for
several final states, where the electron and the positron are created in the process 
$\nu \to \nu e^-_{(n)} e^+_{(\ell)}$ in the lower excited Landau levels.
%
%---------------------------------------------------------------------------------------------
%---------------------------------------------------------------------------------------------
%
The graph shows that for the channels $(n, \ell) = (1,0), (2,0), (3,0)$, and accordingly $(0,1), (0,2), (0,3)$,
the threshold energy is lower than for the channel $(1,1)$. The analysis show that the expressions for
polarization amplitudes~(\ref{eq:M_--_nl}) and~(\ref{eq:M_+-_nl}) are significantly simplified when one of
the particles, electron or positron, is born in the ground Landau level. The amplitudes take the simplest form  
in the case of a sufficiently strong field, $2 \beta n \gg m_e^2$. The amplitudes for the channels
$(n,0)$ and $(0,n)$, where $n \geqslant 1$, and with $s = s' = -1$ can be written with the single formula:
\begin{eqnarray}
{\cal M}^{--}_{n0,0n} & = & \eta \, \frac{G_{\mathrm{F}}\, (C_V - \zeta \, C_A) \, 
\sqrt{2 \beta n}}
{2 \, \sqrt{(n-1)!} } \, \sqrt{q_{\mprl}^2 - 2 \beta n} 
\left( \frac{q_{\mprp}^2}{2 \beta}\right)^{n/2} \, \E^{- q_{\mprp}^2/(4 \beta)}
\nonumber\\[2mm]
& \times &
\left[ \frac{(j \tilde \varphi q) - \zeta \, (j \tilde \Lambda q)}{q^2_{\mprl}}
+ \zeta \, \frac{(j \Lambda q) \mp \I \, (j \varphi q)}{q^2_{\mprp}} \right] \,.
\label{eq:M_--_n0} 
\end{eqnarray}
Two more non-zero polarization amplitudes differ from~(\ref{eq:M_--_n0}) by the phase factors only, 
so that we have
\begin{eqnarray}
&&|{\cal M}^{+-}_{n0}|  =  |{\cal M}^{--}_{n0}| \,,
\label{eq:M_+-_n0}\\
&&|{\cal M}^{-+}_{0n}|  =  |{\cal M}^{--}_{0n}| \,.
\label{eq:M_-+_n0}
\end{eqnarray}
In the considered approximation, the widths of the processes 
$\nu \to \nu e^-_{(n)} e^+_{(0)}$ and $\nu \to \nu e^-_{(0)} e^+_{(n)} \, (n \geqslant 1)$ 
can be written with the single formula as the following integral over the final neutrino 
momentum: 
\begin{eqnarray}
W_{n0,0n} & = & \frac{G_{\mathrm{F}}^2 \, \beta}{(2 \pi)^4 \, (n-1)! \, E} \, 
\int \frac{\D^3 P'}{E'} \; \Theta(q^2_{\mprl} - 2 \, \beta \, n) \,
\frac{- q^2}{(q^2_{\mprl})^2} 
\left( \frac{q_{\mprp}^2}{2 \beta}\right)^{n-1} \, \E^{- q_{\mprp}^2/(2 \beta)}
\nonumber\\[2mm]
& \times &
\bigg \lbrace (C_V^2 + C_A^2)\bigg [\left( P^2_{\mprl} - P'^2_{\mprl}\right)^2 
+ 4 \left( P \, P' \right)^2_{\mprl} 
- 2 \left( P^2_{\mprl} + P'^2_{\mprl}\right) \left( P \, P' \right)_{\mprl}  
\nonumber\\[2mm]
& \mp &
 2 \, q^2_{\mprl} \left( P \tilde \varphi P' \right) \bigg ] 
- 2 \, C_V C_A \left( P^2_{\mprl} - P'^2_{\mprl}\right) [ \, 2 \left( P \tilde \varphi P' \right) \mp q^2_{\mprl} ] 
 \bigg \rbrace ,
\label{eq:W_n0_inv}
\end{eqnarray}
where $\Theta(x)$ is the step function.

It is convenient to perform further integration over the final neutrino momentum, 
without loss of generality, not in an arbitrary reference frame $K$, 
but in the special frame $K_0$, 
where the initial neutrino momentum is perpendicular to the magnetic
field, $P_z = 0$. One can then return from 
$K_0$ to $K$ by the Lorentz transformation along the field
(recall that the field is invariant with respect to this transformation). 
Indeed, one can see that the product $E \; W$, determined from
Eq.~(\ref{eq:W_n0_inv}), contains only the invariants.

In the formula~(\ref{eq:W_n0_inv}), it is convenient to use the dimensionless cylindrical
coordinates in the space of the final neutrino momentum vector $\bs P\,'$:
\begin{eqnarray}
&&\rho = \sqrt{P_x'^2 + P_y'^2}/E_{\mprp} \,, 
\qquad \mbox{tg}\,\phi = P_y'/P_x'\,, 
\qquad z = P_z'/E_{\mprp} \,,
\nonumber\\[2mm]
&&r = E'/E_{\mprp} = \sqrt{\rho^2 + z^2} \,.
\label{eq:variab}
\end{eqnarray}
Here, $E_{\mprp}$ is the energy of the initial neutrino in the frame $K_0$, 
which is connected with its energy $E$ in an arbitrary frame $K$ by the 
relation $E_{\mprp} = E \sin \theta$.  

After the change of variables~(\ref{eq:variab}), the expression of~(\ref{eq:W_n0_inv}) 
can be transformed to
\begin{eqnarray}
E \, W_{n0,0n} & = & \frac{G_{\mathrm{F}}^2 \, \beta \, E_{\mprp}^4}{4 \pi^3 \, (n-1)!}
\left( \frac{E_{\mprp}^2}{2 \beta}\right)^{n-1} 
\int\limits_0^{1-\sqrt{b_n}} \D \rho\, \rho 
\, \int\limits_{-Z_0}^{Z_0} 
\, \frac{\D z}
{r\, (1 - 2 r + \rho^2)^2}
\nonumber\\[2mm]
& \times &
\bigg \lbrace (C_V^2 + C_A^2)\bigg [(1 - \rho^2)^2 + 4 \, r^2 - 2 r (1 + \rho^2) 
\mp 2 (1 - 2 r + \rho^2) z \bigg ] 
\nonumber\\[2mm]
& - &
 2 \, C_V C_A (1 - \rho^2) \bigg [ 2 z \mp (1 - 2 r + \rho^2 ) \bigg ] \bigg \rbrace  
\, \int\limits_0^{2 \pi} \frac{\D \phi}{2 \pi} \, (r - \rho \cos \phi)  
\nonumber\\
& \times &
(1 - 2 \rho \, \cos \phi + \rho^2)^{n-1} \,
\exp \left( -\frac{E_{\mprp}^2}{2 \beta} (1-2\rho\, \cos \phi + \rho^2) \right),
\label{eq:W_n0}
\end{eqnarray}
where 
\begin{eqnarray}
b_n = \frac{2 \beta n}{E_{\mprp}^2}, \qquad
Z_0 = \frac{1}{2} \, \sqrt{\left(1 + \rho^2 - b_n \right)^2 - 4 \rho^2}.
\label{eq:notat}
\end{eqnarray}
The terms in Eq.~(\ref{eq:W_n0}) linearly depending on $z$ determine the asymmetry 
of the final neutrino emission with respect to the magnetic field, and 
these terms do not contribute to the probability. 
However, they appear to be important when the differential probability from Eq.(\ref{eq:W_n0}) 
is used for calculating the asymmetry of the averaged neutrino momentum loss. 

The expression (\ref{eq:W_n0}) can be integrated analytically over the $\phi$ angle, 
e.g. for $n=1$ one obtains:
\begin{eqnarray}
&& E \, W_{10,01} = \frac{G_{\mathrm{F}}^2 \, \beta \, E_{\mprp}^4}{2 \pi^3}
\, \E^{-1/b}
\int\limits_0^{1-\sqrt{b}} \D \rho\, \rho \, \E^{-\rho^2/b}
\, \int\limits_{0}^{Z_0} 
\, \frac{\D z}
{r\, (1 - 2 r + \rho^2)^2}
\nonumber\\[2mm]
&& \times 
\bigg \lbrace (C_V^2 + C_A^2)\bigg [(1 - \rho^2)^2 + 4 \, r^2 - 
 2 r (1 + \rho^2) \bigg ] 
\nonumber\\[2mm]
&& \pm 
 2 \, C_V C_A (1 - \rho^2) \, (1 - 2 r + \rho^2 ) \bigg \rbrace  
\left[ r \, I_0 \! \left( \frac{2 \rho}{b} \right) - \rho \, I_1 \! \left( \frac{2 \rho}{b} \right) \right] ,
\label{eq:W_n0_2}
\end{eqnarray}
where $b = 2 \beta /E_{\mprp}^2$, and $I_n (x)$ are the modified Bessel functions~\cite{Gradshtein}. 

The probabilities of the process $\nu \to \nu e^-_{(n)} e^+_{(\ell)}$,  
evaluated numerically as the functions of the initial neutrino energy and 
on the strength of the magnetic field for all channels considered in 
Ref.~\refcite{Dicus:2007}, where the electron and positron are created 
in the lower Landau levels, are in a good agreement with the results of that paper.

%%%%%%%%%%%%%%%%%%%%%%%%%%%%%%%%%%%%%%%%%%%%%%%%%%%%%%%%%%%%%%%%%%%%%%%%%%%%%%%%
\section{Neutrino energy and momentum losses}	
\label{sec:losses}
%%%%%%%%%%%%%%%%%%%%%%%%%%%%%%%%%%%%%%%%%%%%%%%%%%%%%%%%%%%%%%%%%%%%%%%%%%%%%%%%

The probability of the $\nu \to \nu e^- e^+$ process defines 
its partial contribution into the neutrino opacity of the medium. 
The estimation of the neutrino mean free 
path with respect to this process gives the result which is too large~\cite{KM_Book_2013} 
compared with the typical size of a compact astrophysical object, e.g. the supernova remnant, 
where a strong magnetic field could exist. 
However, a mean free path does not exhaust the neutrino physics in 
a medium. In astrophysical applications, we could consider 
the values that probably are more essential, namely, the mean values 
of the neutrino energy and momentum losses, caused by the influence of an external magnetic field. 
These values can be described by the four-vector of losses $Q^{\alpha}$, 
\begin{equation}
Q^\alpha \, = \, E \int q^\alpha \, \D W = - E \, ({\cal I}, {\bs F}) \,.
\label{eq:Q0}
\end{equation}
where $q$ is the difference of the momenta of the initial and final neutrinos, 
$q = P - P'$, $\D W$ is the total differential probability of the process. 
The zeroth component of $Q^{\alpha}$ is connected with the mean energy lost 
by a neutrino per unit time due to the process considered, 
${\cal I} = \D E/\D t$.
The space components of the four-vector~(\ref{eq:Q0}) are similarly 
connected with the mean neutrino momentum loss per unit time, 
${\bs F} = \D {\bs P}/\D t$. 
It should be noted that the four-vector of losses $Q^{\alpha}$ can be used for evaluating 
the integral effect of neutrinos on plasma 
in the conditions of not very dense plasma, e.g. of a supernova envelope, 
when an one-interaction approximation of a neutrino with plasma 
is valid.
~\cite{Ruffert:1997,Kuznetsov:1997a,Kuznetsov:1997b,Birkl:2007}
Otherwise, to evaluate the neutrino energy deposition in the conditions of dense medium, 
e.g. in the supernova core, where a neutrino participates in multiple interactions, 
one should solve general neutrino transport 
equations.~\cite{Mezzacappa:2005,Woosley:2005,Kotake:2006,Janka:2012} 

In Ref.~\refcite{Beloborodov:2011}, the formula (10) for the energy deposition rate was taken, which 
was calculated in the crossed field limit.~\cite{Kuznetsov:1997a,Kuznetsov:1997b} By the way, the value $q^\alpha$ 
defined by Eq.~(10) of Ref.~\refcite{Beloborodov:2011} is not the 4-vector while the value 
$Q^\alpha = E \, q^\alpha$ is. 
However, in the region of the physical parameters 
used in Ref.~\refcite{Beloborodov:2011} ($B$ to 180 $B_e$, $E_\nu$ to 25 MeV), the approximation of a crossed field is poorly applicable,
as well as the approximation of a superstrong field when $e^- e^+$ are created in the ground Landau level.
The contribution of the next Landau levels which can be also excited, should be taken into account.
We present here the results of our calculation of the mean neutrino energy losses caused by
the process $\nu \to \nu e^- e^+$ in a moderately strong magnetic field, i.e. in the conditions 
of the Kerr black hole accretion disk.  

We parametrize the energy deposition rate as:
\begin{equation}
Q_0 \, = \, (C_V^2 + C_A^2) \, \sigma_0 \, m_e^4\,E \; f \! \left( \frac{E}{m_e}\,, \frac{B}{B_e} \right) \,, 
\label{eq:Q0_2}
\end{equation}
where 
$\sigma_0 = 4 \, G_{\mathrm{F}}^2 \, m_e^2/\pi$, and the dependence on the initial neutrino energy and the field 
strength is described by the function $f ( {E}/{m_e}\,, {B}/{B_e} )$.
This function calculated in Refs.~\refcite{Kuznetsov:1997a,Kuznetsov:1997b} in the crossed field limit had the form
\begin{equation}
f^{(cr)} (y, \eta) \, = \, \frac{7\, y^2 \, \eta^2 }{1728 \, \pi^2} \, \ln (y \, \eta) \,, 
\label{eq:f_(cr)}
\end{equation}

On the other hand, in the strong field limit when both electron and positron are born in the ground Landau level, 
the function $f (y, \eta)$ was also calculated in Refs.~\refcite{Kuznetsov:1997a,Kuznetsov:1997b} and can be presented in the form
\begin{equation}
f^{(00)} (y, \eta) \, = \, \frac{\eta \, y^4}{32 \, \pi^2} \, 
\int\limits_0^1 \D \rho\, \rho (1 - \rho^2)^2 \,
\exp \! \left( - \frac{y^2 (1 + \rho^2)}{2 \eta} \right)
I_0 \! \left( \frac{y^2}{\eta} \, \rho \right).
\label{eq:f_(00)}
\end{equation}

In conditions of moderately strong magnetic field, when the electron and the positron are created in the process
$\nu \to \nu e^-_{(n)} e^+_{(\ell)}$ in the $n$th and $\ell$th Landau levels, the result has more complicated form. 
It is significantly simplified when one of
the particles, electron or positron, is born in the ground Landau level. 
We obtain the contribution of the channels $\nu \to \nu e^-_{(n)} e^+_{(0)}$ and $\nu \to \nu e^-_{(0)} e^+_{(n)}$ to the function $f (y, \eta)$ as follows:
\begin{eqnarray}
&&f^{(n0+0n)} (y, \eta) \, = \, \frac{\eta \, y^4}{4 \pi^2 (n-1)!} \, 
\left( \frac{y^2}{2 \eta} \right)^{n-1}
\int\limits_0^{1-\sqrt{b_n}} \D \rho\, \rho \, \int\limits_0^{Z_0} \frac{\D z (1-r)}{r (1-2 r + \rho^2)^2} 
\nonumber\\[2mm]
&& \times 
\left[(1 - \rho^2)^2 + 4 r^2-2 r (1+\rho^2) \right] 
\int\limits_0^{2 \pi} \frac{\D \phi}{2 \pi} (r-\rho \cos \phi) 
\nonumber\\[2mm]
&& \times \,
 (1 - 2 \rho \cos \phi + \rho^2)^{n-1} 
\exp \! \left( - \frac{y^2 (1 - 2 \rho \cos \phi + \rho^2)}{2 \eta} \right).
\label{eq:f_(0n)}
\end{eqnarray}

\begin{figure}[!t] 
\centering 
\includegraphics*[width=\textwidth]{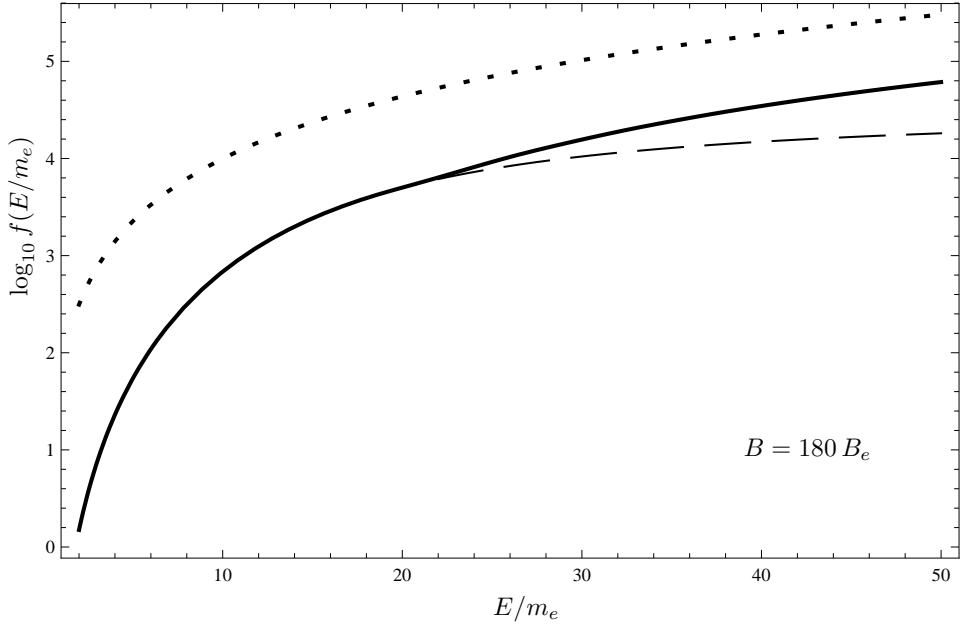}
\caption{The function $f (E/m_e)$ for $B=180 \,B_e$ obtained in the crossed field limit (dotted line), 
with $e^- e^+$ created at the ground (0,0) Landau level (dashed line), and for the sum of all lower 
Landau levels which are excited in this energy interval according 
to the condition~(\ref{eq:condE})(solid line).}
\label{fig:function180}
\end{figure}

\begin{figure}[ht] 
\centering 
\includegraphics*[width=\textwidth]{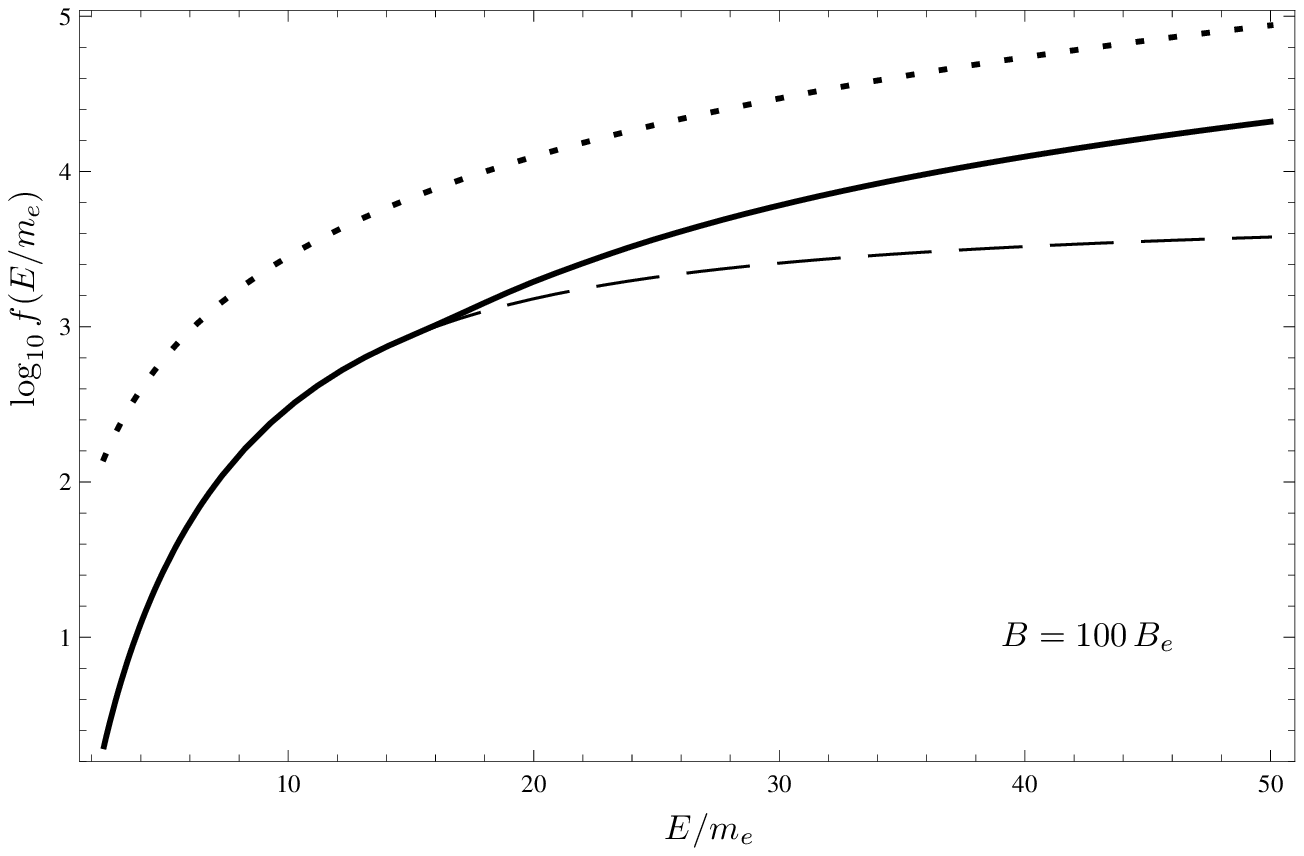}
\caption{The same as in Fig.~\ref{fig:function180}, for $B=100 \,B_e$.}
\label{fig:function100}
\end{figure}

\begin{figure}[!t] 
\centering 
\includegraphics*[width=\textwidth]{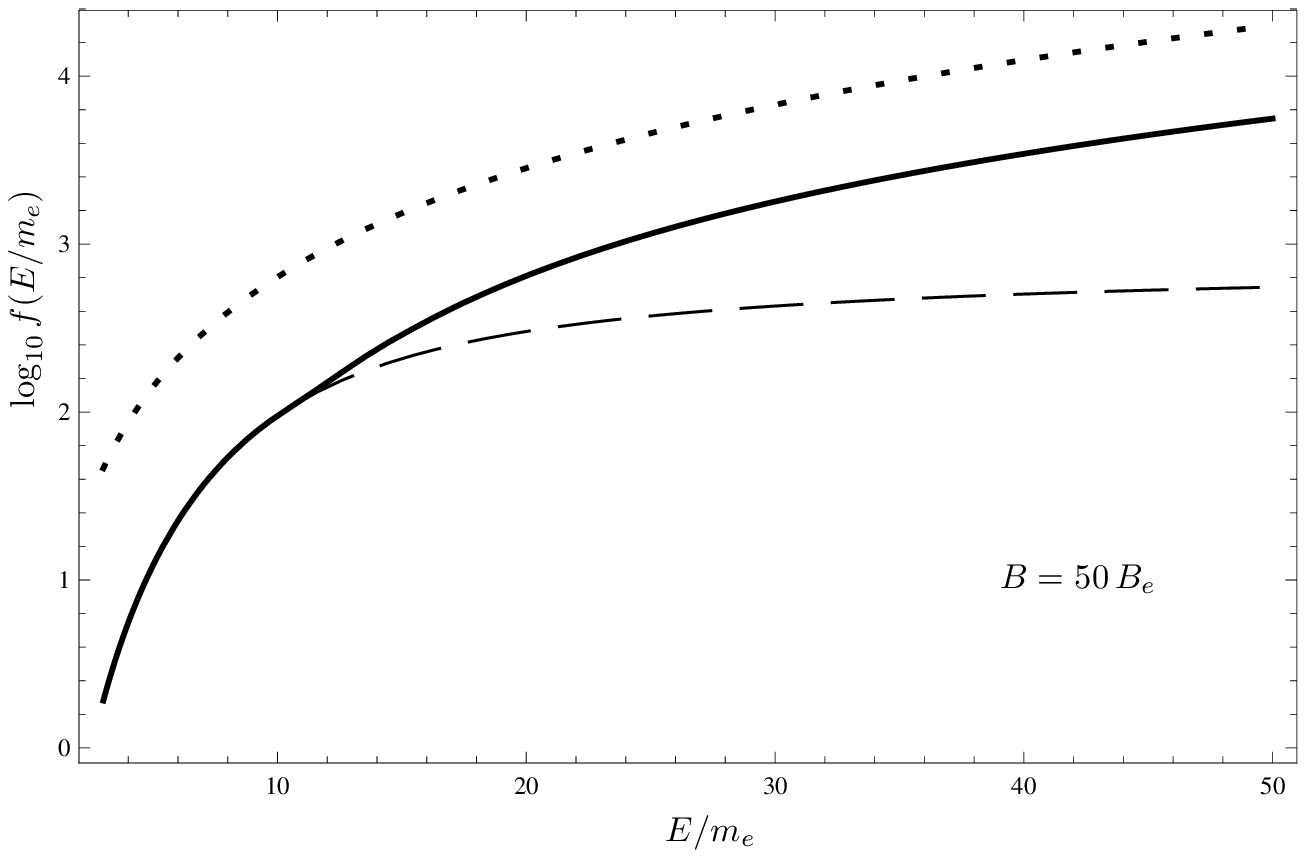}
\caption{The same as in Fig.~\ref{fig:function180}, for $B=50 \,B_e$.}
\label{fig:function50}
\end{figure}

In Figs.~\ref{fig:function180}--\ref{fig:function50}, the function $f (y, \eta)$ obtained in different approximations 
is shown at $B = 180 B_e, \,100 B_e, \,50 B_e$. 
It can be seen that the crossed field limit gives the overstated result which is in orders of magnitude greater 
than the sum of the contributions of lower excited Landau levels.   
On the other hand, the results with $e^- e^+$ created at the ground Landau level give the main contribution to 
the energy deposition rate, and almost exhaust it at $B = 180 B_e$. 

This would mean that the conclusion~\cite{Beloborodov:2011} that the contribution of the process 
$\nu \to \nu e^- e^+$ to the efficiency of the electron-positron plasma production by neutrino exceeds
the contribution of the annihilation channel $\nu \bar\nu \to e^- e^+$, and that the first process 
dominates the energy deposition rate, does not have a sufficient basis.  
A new analysis of the efficiency of energy deposition by neutrinos through both processes, $\nu \bar\nu \to e^- e^+$ 
and $\nu \to \nu e^- e^+$, in a hyper-accretion disc around a black hole should be performed, with taking 
account of our results for the process $\nu \to \nu e^- e^+$ presented here. 

%---------------------------------------------------------------------------------------------
%---------------------------------------------------------------------------------------------

%---------------------------------------------------------------------------------------------
%---------------------------------------------------------------------------------------------

%---------------------------------------------------------------------------------------------
%---------------------------------------------------------------------------------------------

%%%%%%%%%%%%%%%%%%%%%%%%%%%%%%%%%%%%%%%%%%%%%%%%%%%%%%%%%%%%%%%%%%%%%%%%%%%%%%%%
\section{Conclusions}	
\label{sec:Conclusions}
%%%%%%%%%%%%%%%%%%%%%%%%%%%%%%%%%%%%%%%%%%%%%%%%%%%%%%%%%%%%%%%%%%%%%%%%%%%%%%%%

In the paper, a calculation is performed of the mean value of the neutrino energy loss due to the process 
of electron-positron pair production, 
$\nu \to \nu e^- e^+$, in the magnetic field of an arbitrary strength at which the electrons and positrons 
can be produced in the states corresponding to the excited Landau levels, which could be essential
in astrophysical applications. In calculations, the exact solutions 
were used of the Dirac equation for an electron in a magnetic field, which are simultaneously the eigenfunctions 
of the covariant operator of a magnetic polarization. This allowed to calculate the partial amplitudes 
corresponding to different polarization states of the electrons and positrons in a manifestly relativistic 
invariant form. 
The results obtained should be used for calculations of the efficiency of the electron-positron plasma production 
by neutrinos in the conditions of the Kerr black hole accretion disk, regarded by experts as the most likely source 
of a short cosmological gamma-ray burst. In those conditions, the crossed field limit used in the previous 
calculations led to the overstated result which was in orders of magnitude greater 
than the sum of the lower Landau levels. 
The study may be also useful for further development of computational techniques for the analysis 
of quantum processes in an external active environment, particularly in conditions of moderately strong magnetic field, 
when the allowance for the contribution of only the ground Landau level is insufficient.

%%%%%%%%%%%%%%%%%%%%%%%%%%%%%%%%%%%%%%%%%%%%%%%%%%%%%%%%%%%%%%%%%%%%%%%%%%%%%%%%
\section*{Acknowledgments}
%%%%%%%%%%%%%%%%%%%%%%%%%%%%%%%%%%%%%%%%%%%%%%%%%%%%%%%%%%%%%%%%%%%%%%%%%%%%%%%%

We dedicate this paper to the blessed memory of our teacher, colleague and friend 
Nickolay Vladimirovich Mikheev who passed away on June 19, 2014. 

The study was performed with the support by the Project No.~92 within the base part of the State Assignment 
for the Yaroslavl University Scientific Research, and was supported in part by the 
Russian Foundation for Basic Research (Project No. \mbox{14-02-00233-a}).

%%%%%%%%%%%%%%%%%%%%%%%%%%%%%%%%%%%%%%%%%%%%%%%%%%%%%%%%%%%%%%%%%%%%%%%%%%%%%%%%
%\section*{References}
%%%%%%%%%%%%%%%%%%%%%%%%%%%%%%%%%%%%%%%%%%%%%%%%%%%%%%%%%%%%%%%%%%%%%%%%%%%%%%%%

%\begin{thebibliography}{000} %for 3 digits

\end{document}